\documentclass[prl,aps,showpacs,twocolumn,eqsecnum,a4paper,ltxgrid]{revtex4}

\usepackage{graphicx}
\usepackage[sort&compress]{natbib}
\usepackage{amssymb}
\usepackage{booktabs}
\usepackage[utf8]{inputenc}
\usepackage[english]{babel}
\usepackage{amsfonts}
\usepackage{amsmath}
\usepackage{array}
\usepackage{multirow}
\usepackage{color}
\usepackage{subdepth}
\usepackage{babel}
\usepackage{todonotes}
\usepackage{chngcntr}
\usepackage{epsfig}

\makeatletter

\newcommand{\mKet}[1]{| #1 \rangle}

\newcommand{\mAve}[1]{\left \langle #1 \right \rangle}
\newcommand{\Eq}[1]{Eq.\,(\ref{#1})}
\newcommand{\Eqs}[1]{Eqs.\,({#1})}

\newcommand{\FT}[1]{\check{#1}}
\newcommand{\e}{\mathrm{e}}
\renewcommand{\i}{\mathrm{i}}
\newcommand{\diff}{\mathrm{d}}
\newcommand{\Fig}[1]{Fig.\,\ref{#1}}

\newcommand{\cm}{cm$^{-1}$}
\newcommand{\SI}{Supplement}

\@ifundefined{textcolor}{}
{%
 \definecolor{BLACK}{gray}{0}
 \definecolor{WHITE}{gray}{1}
 \definecolor{RED}{rgb}{1,0,0}
 \definecolor{GREEN}{rgb}{0,1,0}
 \definecolor{BLUE}{rgb}{0,0,1}
 \definecolor{CYAN}{cmyk}{1,0,0,0}
 \definecolor{MAGENTA}{cmyk}{0,1,0,0}
 \definecolor{YELLOW}{cmyk}{0,0,1,0}
}

\usepackage{blindtext}
\makeatother

\counterwithout{equation}{section}

\begin{document}

\title{Nuclear dynamical correlation effects in X-ray spectroscopy from a time-domain perspective}

\author{Sven Karsten$^1$}
\author{Sergei D. Ivanov$^1$}
\email{sergei.ivanov@uni-rostock.de}
\author{Saadullah G. Aziz$^2$}
\author{Sergey I.\ Bokarev$^1$}
\email{sergey.bokarev@uni-rostock.de}
\author{Oliver K\"uhn$^1$}
\affiliation{$^1$Institute of Physics, University of Rostock, Albert-Einstein-Str. 23-24, 18059 Rostock, Germany}
\affiliation{$^2$Chemistry Department, Faculty of Science, King Abdulaziz University, 21589  Jeddah, Saudi Arabia}

\date{\today}

\begin{abstract}
To date X-ray spectroscopy has become a routine tool that can reveal highly local and element-specific information on the electronic structure of atoms in complex environments.
%
%
Here, we focus on nuclear dynamical effects in X-ray spectra and develop a rigorous time-correlation method 
employing ground state molecular dynamics simulations.
%
The importance of nuclear correlation phenomena is demonstrated by comparison against the results from the conventional sampling approach for gas phase water.
In contrast to the first-order absorption, second-order resonant inelastic scattering spectra 
exhibit pronounced fingerprints of nuclear motions.
%
The developed methodology does not depend on the accompanying electronic structure method in principle as well as on the spectral range and, thus, can be applied to, e.g., UV and X-ray photo-electron and Auger spectroscopies.

\end{abstract}

\pacs{
33.20.Rm, 
87.10.Tf, 
34.50.Gb, 
31.15.A- 
}

\maketitle

\paragraph{Introduction.}
Constant increase of spectral resolution and rapid development of various spectroscopies, covering broad energy ranges from radio frequencies to extra hard radiation, opens new horizons for a molecular scientist to investigate more and more intricate and delicate phenomena.
When it comes to obtaining highly local and element-specific information on the electronic structure,
X-ray spectrosopies stand out~\cite{stohr2013}.
Popular variants include first order X-ray absorption spectra (XAS) and second order resonant inelastic X-ray scattering (RIXS) techniques.
The former focuses on the electronic transitions where a core electron is excited to the manifold of unoccupied molecular orbitals (MOs), whereas the latter detects the emission signal resulting from the refill of a core-hole by electrons occupying valence MOs.
%
Although X-ray spectroscopy usually targets electronic transitions, the vibrational ones as well as the accompanying nuclear dynamics have recently received growing attention~\cite{Hennies2005, Ljungberg2011,Rubensson2013, Guillemin2013,Dong2013,Bohinc2013, Pietzsch2011, Hennies2010}.
%
Remarkably, the RIXS spectra of liquid water and alcohols initiated active ongoing debates in the last decade~\cite{Lange2013, Schreck2014,Fransson-CR-2016}, with controversial interpretations, among others involving different aspects of nuclear dynamics~\cite{Sellberg2015}.

Conventionally, electronic spectra are obtained via single point electronic structure calculations combined with models such as the multi-mode Brownian oscillator one to include broadening on phenomenological level ~\cite{Mukamel-Book}.
A big step forward is to sample nuclear distributions in the phase space via molecular dynamics (MD) methods~\cite{Kuehn-Book, marx2009, Ivanov-PCCP-2013},
leading to a more realistic description of conformational and environmental effects~\cite{Sun2011, Jena2015, Weinhardt2015, Leetmaa2010}, although lacking information about correlated nuclear motion.
Here, we propose an approach to theoretical calculations of XAS and RIXS spectra based on time-correlation functions obtained from the time evolution provided by electronic 
ground state MD simulations, analogous to infrared and UV/Vis spectroscopies~\cite{Heller1978,Heller-JCP-1979,lawrence2002,harder2005,Ivanov-PCCP-2013,Mukamel-Book}.

The central message of this Letter is that nuclear correlation effects are essential for X-ray spectroscopy.
We exemplify this on oxygen K-edge spectra of a typical and highly relevant system:
gas phase water, by comparison against the results of the aforementioned sampling approach. 
Importantly, only RIXS, being a second order process, appears sensitive to them, much like non-linear optical spectra provide more detailed insight into the underlying dynamical processes~\cite{Kuehn-Book,Mukamel-Book}.

\paragraph{Theory.}

The proposed method is based on the rigorous derivation from the first principles employing the interaction representation picture and the dynamical classical limit~\cite{Kuehn-Book,Mukamel-Book}.
Technically, the time dependence of all quantities is provided by the classical MD in the electronic ground state.
The quantities themselves are obtained from an approximate solution of the time-independent electronic Schr\"odinger equation at each time instance.
The working expressions for XAS, $\mathcal{X}(\Omega)$, and RIXS, $\mathcal{R}(\Omega, \omega)$,  amplitudes can be derived starting from the linear and Raman third-order response functions~\cite{Mukamel-Book}, correspondingly, see \SI.
%
Note that from the practical standpoint it is more convenient to work in frequency domain.

\begin{figure}
\includegraphics[scale=0.4]{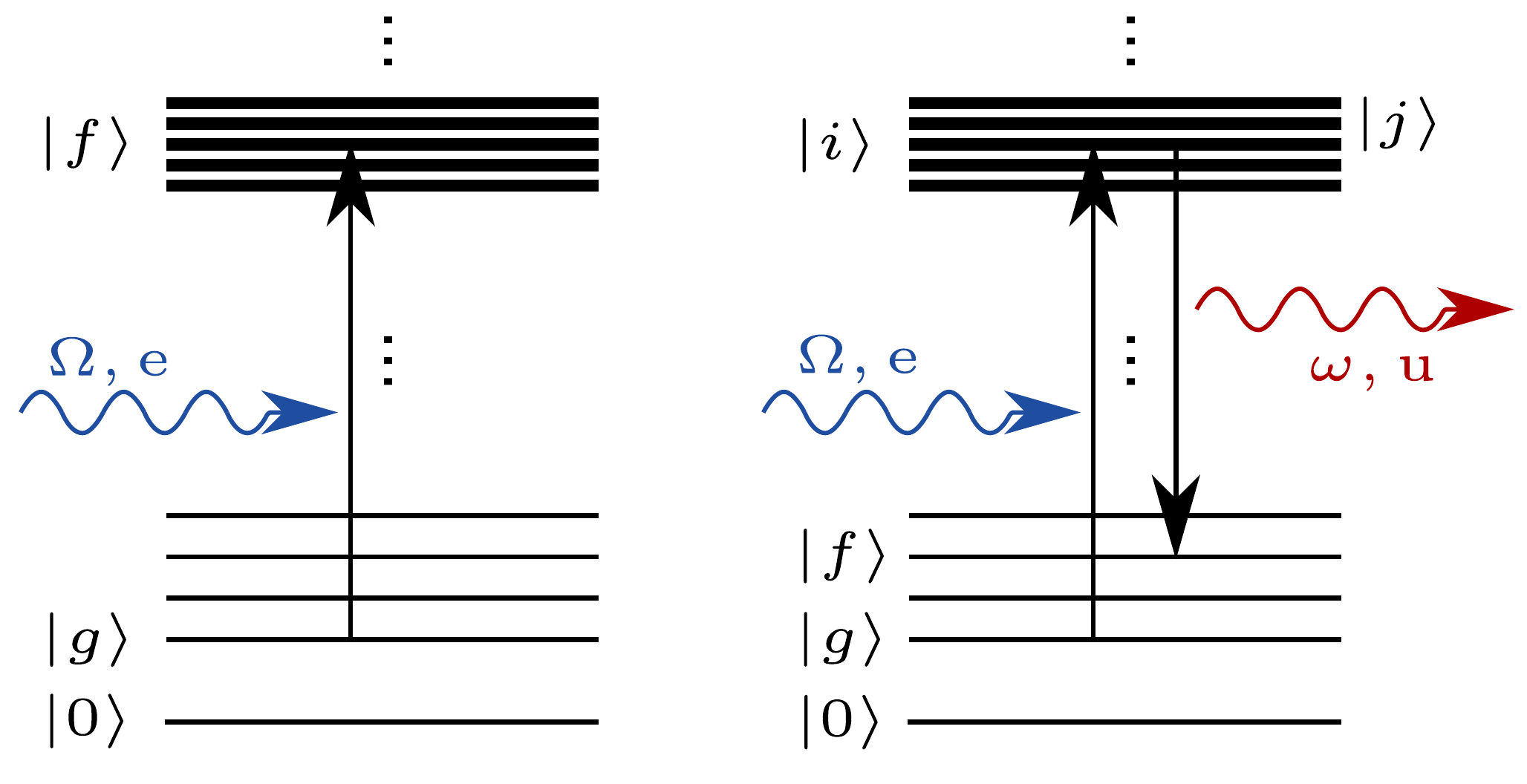}
\caption{\label{fig:sketch}
The schematic sketch of XAS (left) and RIXS (right) processes, see text.
}
\end{figure}

The XAS process consists of exciting the system from an initial state $\mKet{g}$ to a final core-excited state $\mKet{f}$ by absorbing light with angular frequency $\Omega$ and polarisation $\mathbf{e}$, see left panel in \Fig{fig:sketch}.
%
Similarly, in RIXS the system is first excited to a core-excited intermediate state $\mKet{i}$ or $\mKet{j}$ 
and then transits to the valence final state $\mKet{f}$ by emitting light with the frequency $\omega$ and polarisation $\mathbf{u}$, see right panel therein.
%
The final expressions for the spectral amplitudes in atomic units read
\begin{equation}
\label{eq:XAS_final_Fourier}
\mathcal{X}(\Omega)= \frac{1}{T} \sum\limits_{g,f} \mAve{ \FT{\bar{M}}^{g}_f(\bar{\omega}_{fg}-\Omega) \FT{\mathcal{\bar{M}}}^{f}_g(\Omega-\bar{\omega}_{fg})}
\enspace 
\end{equation}
and
\begin{align}
\label{eq:RIXS_final_Fourier}
&\mathcal{R}(\Omega,\omega)=\frac{1}{T} \sum\limits_{g,f, i, j}  \nonumber \\
& \left \langle \intop_{-\infty}^{\infty} \right. \diff \omega_1  \FT{\bar{M}}^{g}_j\left(\bar{\omega}_{jg}-[\Omega-\omega_1 ]\right) \FT{\Delta}_j(\omega_1 ) \FT{\bar{M}}^{j}_f\left([\omega -\omega_1] - \bar{\omega}_{jf}\right)  \nonumber \\
& \left. \intop_{-\infty}^{\infty} \diff \omega_2  \FT{\bar{M}}^{f}_i\left(\bar{\omega}_{if}-[\omega  - \omega_2] \right) \FT{\Delta}_i(-\omega_2 ) \FT{\mathcal{\bar{M}}}^{i}_g\left([\Omega -\omega_2] - \bar{\omega}_{ig}\right)  \right \rangle
\enspace .
\end{align}
Here, inversed hats denote the Fourier-transformed quantities and the ``dressed'' transition dipole moments read
\begin{eqnarray}
\bar{M}^{f}_g(t,0)& := & D^{f}_{g}(t) \exp{\left[ \i \int_{0}^{t} \diff \tau U_{fg}(\tau)\right]} 
\nonumber\\
\mathcal{\bar{M}}^{f}_g(t,0)& := &\mathcal{W}_g(t)D^{f}_{g}(t) \exp{\left[ \i \int_{0}^{t} \diff \tau U_{fg}(\tau)\right]}
\label{eq:dressed_dipole}
\enspace,
\end{eqnarray}
%
where $D_g^f$
are the transition dipole moments from $\mKet{g}$ to $\mKet{f}$ time-evolved with respect to the Hamilton function of the electronic ground state.
The gap fluctuation reads $U_{fg}(\tau):=\Delta E_{fg}(\tau)-\bar{\omega}_{fg}$, where $\Delta E_{fg}$ is the electronic energy gap and
$\bar{\omega}_{fg}:=1/T\int_{0}^{T} \diff \tau\Delta E_{fg}(\tau)$
is the mean transition frequency averaged over a trajectory of length $T$.
Further $\FT{\Delta}_j(\omega ' ):=1/(2\pi)\sqrt{\Gamma_j/\pi}(\Gamma_j+\i \omega ')^{-1}$ is the damping function in frequency domain, with $\Gamma_j$ being the lifetime broadening of the intermediate state $\mKet{j}$.
The lifetimes are thus allowed for by means of a simple phenomenological model accounting for the Auger decay.
%
Additionally, the spectra are convoluted with the Gaussian of width $\sigma$ along the $\Omega$-axis, responsible for the bandwidth of the excitation pulse.
Finally, the weighting function
\begin{equation}
\label{eq:weighting_factor}
 \mathcal{W}_g(t):=\e^{-\Delta E_{g0}(t)/kT}/ \mAve{ \sum_g\e^{-\Delta E_{g0}(0)/kT}} 
\enspace,
\end{equation}
with $\langle \ldots\rangle$ here and in \Eqs{\ref{eq:XAS_final_Fourier},\ref{eq:RIXS_final_Fourier}} standing for the classical canonical average with respect to the Hamilton function of the ground state, $H_0$, see \Fig{fig:sketch}.
Note that for the present case of gas phase water, there is only one initial state $\mKet{g}$ that coincides with the ground state $\mKet{0}$.
We note in passing that \Eqs{\ref{eq:XAS_final_Fourier},\ref{eq:RIXS_final_Fourier}} can be derived starting from the Fermi's Golden rule and the Kramers-Heisenberg expression, respectively.

\begin{figure}[tb]
  \includegraphics[width=0.99\columnwidth]{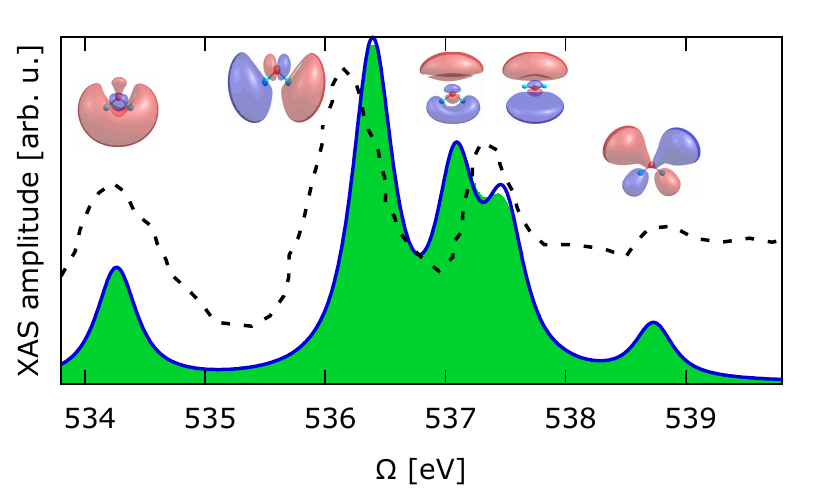}
\caption{\label{fig:XAS}
XAS amplitudes for gas phase water.
Dashed line represents the respective experimental data from Ref.~\citenum{Lange2012}.
Blue line depicts the time-correlation approach results according to \Eq{eq:XAS_final_Fourier}, whereas filled green curve corresponds to the sampling method.
The unoccupied MOs to which the transition is performed are shown near the respective spectral peaks.
}
\end{figure}

\paragraph{Computational details.}
The MD simulations have been performed using \textsc{Gromacs} ver.~4.6.5~\cite{GROMACS} employing the anharmonic qSPC/Fw water model with a Morse O--H potential~\cite{Paesani-JCP-2006}.
A set of $140$ uncorrelated initial conditions has been sampled from an $NVT$ MD run at $300$\,K further serving as
starting points for $NVE$ trajectories.
The trajectories have been $0.5$\,ps long with a timestep of $0.5$\,fs, yielding a spectral resolution of $\approx\!8\,$meV.
The electronic Schr\"odinger equation for each MD snapshot has been solved via ground state density functional theory with the PBE
functional~\cite{PBE_1996} using ORCA ver.~3.0.3~\cite{orca_2012}.
The def2-QZVPP basis set~\cite{def2_2005} together with (5s5p)/[1s1p] generally contracted Rydberg functions on oxygen have been used.
%
Such a small Rydberg basis does not allow one to reproduce the high-energy tail of the absorption spectrum~\cite{naeslund_JPCA_2003}, but enables the description of the lowest states just above the core-excitation threshold.
The energies of the valence and core-excited states have been approximated by the differences of the respective Kohn-Sham orbital energies; the corresponding dipole transition moments have been calculated with respect to these orbitals~\cite{lee_JACS_2010}, which is known to yield a reasonable compromise between accuracy and efficiency~\cite{Hennies2005,naeslund_JPCA_2003,lee_JACS_2010,Lassalle-Kaiser2013,Pollock2013}.
To preserve the continuous time evolution of the dressed dipoles, the entire manifold of relevant electronic levels has been traced along the MD trajectories in a fully-automated manner.
%
The excitation Gaussian linewidth and the uniform Lorentzian lifetime broadening have been chosen as $\sigma=0.05\,$eV and $\Gamma=0.25\,$fs$^{-1}$, respectively.
The data have been averaged over the molecular orientations assuming the orthogonality of $\mathbf{e}$ and $\mathbf{u}$,
which corresponds to a typical experimental setup,
and the spectra have been shifted {\it globally} by $24.8\,$eV such that the peak structure roughly matches the experimental data~\cite{Lange2012}.
Note that both the sampling and time-correlation approaches employ the same datasets for the consistency of comparison.
For further computational details see \SI.

\paragraph{Results.} 
The fingerprints of nuclear correlations are revealed by comparing the spectra obtained via the sampling and correlation approaches.
%
The XAS amplitudes, \Fig{fig:XAS}, are in fairly good agreement with the experimental data~\cite{Lange2012} shown with the dashed line.
We would like to stress that we do not aim at quantitative reproducing and analyzing experimental data with the rather simple but realistic model employed.
Instead, the focus is on the nuclear correlation effects, manifesting themselves as the differences between the results of the two aforementioned approaches.
%
The first two absorption peaks correspond to the $1\mathrm{s_O}\rightarrow \sigma^*(2\mathrm{s})$ and $1\mathrm{s_O}\rightarrow \sigma^*(2\mathrm{p})$ transitions, whereas the other three stem from the Rydberg $\rm 1s_O\rightarrow 3p_O$ ones,
as is illustrated by the target unoccupied MOs displayed near the respective spectral peaks therein.
Apparently, XAS amplitudes feature only subtle differences in intensities between the sampling and the correlation approach.
%
This illustrates the fact that XAS is not a very sensitive observable for nuclear correlation effects. 

\begin{figure}[tb]
  \includegraphics[width=0.99\columnwidth]{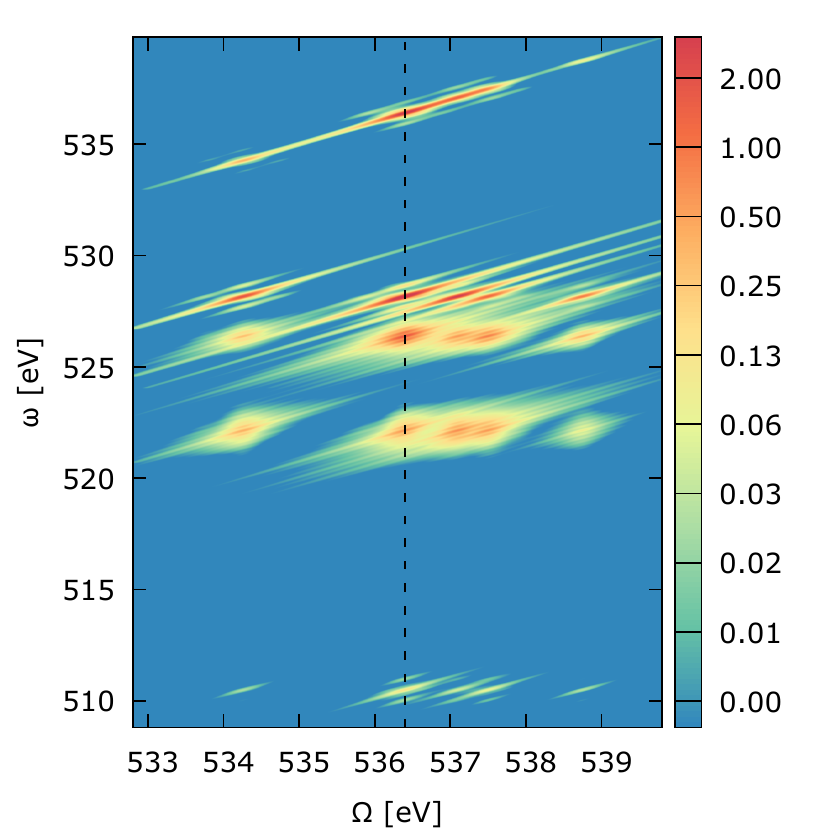}
\caption{\label{fig:RIXS_2d}
2D RIXS spectrum, $\mathcal{R}(\Omega, \omega)$, of gas phase water obtained by means of the time-correlation approach, \Eq{eq:RIXS_final_Fourier}; note the log-scale for intensities depicted with color.
The dashed vertical line indicates the position of the cut depicted in \Fig{fig:RIXS_cut}.
}
\end{figure}

\begin{figure}[tb]
 \begin{center}
  \includegraphics{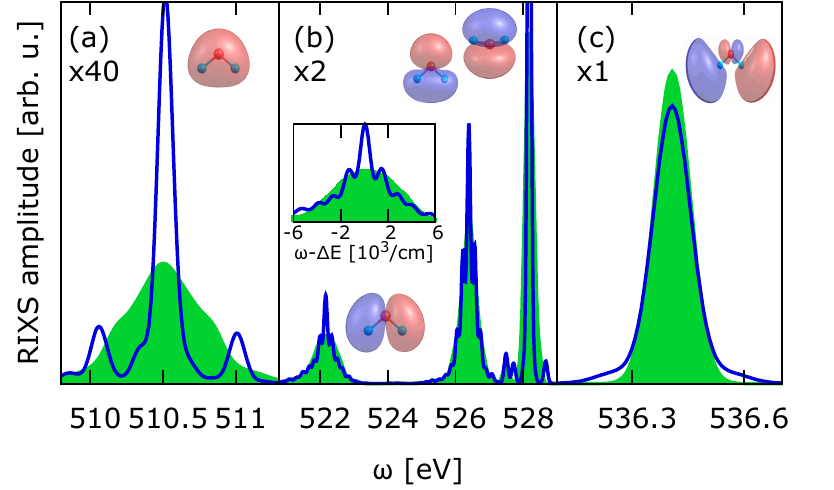}
\caption{\label{fig:RIXS_cut}
A cut through RIXS spectrum in \Fig{fig:RIXS_2d} at $\Omega=536.4$\,eV.
The color-code is the same as in \Fig{fig:XAS}.
Panels depict three relevant spectral ranges.
%
Inset zooms on the left peak in panel (b) that corresponds to $\rm \sigma(2p)\rightarrow1s_O$ transition with $\Delta E=522.2$\,eV.
}
 \end{center}
\end{figure}

Figure~\ref{fig:RIXS_2d} shows a 2D RIXS spectrum obtained via the correlation approach, \Eq{eq:RIXS_final_Fourier}.
Although it gives an overall impression about the spectral shape,
it is hard to make quantitative analysis on its basis.
Therefore, we consider a particular cut for a fixed excitation frequency $\Omega=536.4$\,eV that corresponds to the $1\mathrm{s}_{\rm O}\rightarrow \sigma^*(2\mathrm{p})$ XAS transition, see \Fig{fig:XAS} and vertical line in \Fig{fig:RIXS_2d}.
Three spectral ranges shown in \Fig{fig:RIXS_cut} contain peaks related to transitions from the intermediate (core-excited) states to final (ground or valence-excited) ones, see the respective orbitals from which the core-hole refill takes place.
%
%

Most importantly, the RIXS spectrum obtained via the time-correlation approach exhibit clear traces of nuclear dynamical effects as compared to the sampling one.
First, the two spectra possess notably different lineshapes. 
For instance, the vibronic structure, which is not present in the sampling spectra by construction, is clearly visible for inelastic features shown in panels (a) and (b) in \Fig{fig:RIXS_cut}.
In particular the sidebands for peaks at 510.5\,eV and 528.2\,eV correspond to the O--H stretching mode with the frequency $3800\,$\cm$\,\approx\!0.47$\,eV.
Further, the electronic transitions at $522.2\,$eV (see inset) and 526.4\,eV are coupled to the bending vibrational mode having the frequency of $\approx \!1500$\,\cm.
%
%

Second, the sampling approach exhibits higher intensity of the elastic peak (panel (c)) and lower intensities of the inelastic ones with respect to the correlation method although both techniques employ the same statistics.
The origin of the differences in intensities can be mainly traced back to the complex exponential of the gap fluctuations in the dressed transition dipoles, see \Eq{eq:dressed_dipole}, as will be shown in detail elsewhere.
%
We believe that this makes RIXS spectra more sensitive to nuclear dynamical correlation effects, since the expression in \Eq{eq:RIXS_final_Fourier} contains energy gap fluctuations between \textit{different} pairs of electronic states. 
In contrast, the absorption spectrum, \Eq{eq:XAS_final_Fourier}, depends only on the initial-final gap fluctuations.
%
%


\paragraph{Conclusions and Outlook.}
The simulation protocol allowing for nuclear dynamical phenomena in X-ray spectra has been developed.
This rigorously derived method intrinsically exploits molecular dynamics in the electronic ground state together with a phenomenological dephasing model for core-excited states.
As a word of caution, using the latter model leaves cases that exhibit intricate large-amplitude dynamics in the excited state, e.g., ultrafast dissociation~\cite{Harada-PRL-2013}, outside reach.
Still, this technique does provide an improvement to the description of nuclear dynamical effects in X-ray spectra.
Importantly, these effects have been demonstrated to be essential for X-ray spectroscopy via the comparison against the conventional sampling approach results for gas phase water.
%
Especially RIXS, being a two-photon process, has turned out to be a sensitive technique for the effects in question.
In contrast, XAS, being a one-photon process,
exhibits no traces of the underlying nuclear dynamics.
%
Interestingly,
static (sampling) and dynamic (correlation) nuclear phenomena have been disentangled from each other experimentally, employing RIXS with 
excitation pulses strongly detuned from the resonance~\cite{Schreck2014}.
Thus, a theoretical prediction of fine nuclear effects is expected to stimulate respective high-resolution experiments.

Remarkably, the developed methodology is rather universal and does not conceptually depend on the accompanying electronic structure method.
Further, a similar strategy can be applied to the related photon-in/electron-out techniques, such as photo-electron and Auger spectroscopies as well as to other spectral ranges, e.g., UV/Vis.
We believe that these developments are especially important in view of the recently suggested non-linear X-ray techniques~\cite{Biggs2012,Mukamel2013} that are foreseen to be even more informative and sensitive than the conventional RIXS approach.


\paragraph{Acknowledgement.}
We acknowledge financial support by the Deanship of Scientific Research (DSR), King Abdulaziz University, Jeddah, grant No.\ D-003-435 (S.I.B., O.K.) and the Deutsche Forschungsgemeinschaft KU~952/10-1 (S.K., O.K.), IV~171/2-1 (S.D.I.).
Special thanks go to Fabian Gottwald for technical assistance with the MD simulations of water.

\bibliography{RIXS-add,WaterSolutions}

\begin{thebibliography}{38}
\expandafter\ifx\csname natexlab\endcsname\relax\def\natexlab#1{#1}\fi
\expandafter\ifx\csname bibnamefont\endcsname\relax
  \def\bibnamefont#1{#1}\fi
\expandafter\ifx\csname bibfnamefont\endcsname\relax
  \def\bibfnamefont#1{#1}\fi
\expandafter\ifx\csname citenamefont\endcsname\relax
  \def\citenamefont#1{#1}\fi
\expandafter\ifx\csname url\endcsname\relax
  \def\url#1{\texttt{#1}}\fi
\expandafter\ifx\csname urlprefix\endcsname\relax\def\urlprefix{URL }\fi
\providecommand{\bibinfo}[2]{#2}
\providecommand{\eprint}[2][]{\url{#2}}

\bibitem[{\citenamefont{St{\"o}hr}(2013)}]{stohr2013}
\bibinfo{author}{\bibfnamefont{J.}~\bibnamefont{St{\"o}hr}},
  \emph{\bibinfo{title}{NEXAFS spectroscopy}}, vol.~\bibinfo{volume}{25}
  (\bibinfo{publisher}{Springer Science \& Business Media},
  \bibinfo{year}{2013}).

\bibitem[{\citenamefont{Hennies et~al.}(2005)\citenamefont{Hennies, Polyutov,
  Minkov, Pietzsch, Nagasono, Gel’mukhanov, Triguero, Piancastelli, Wurth,
  {\AA}gren et~al.}}]{Hennies2005}
\bibinfo{author}{\bibfnamefont{F.}~\bibnamefont{Hennies}},
  \bibinfo{author}{\bibfnamefont{S.}~\bibnamefont{Polyutov}},
  \bibinfo{author}{\bibfnamefont{I.}~\bibnamefont{Minkov}},
  \bibinfo{author}{\bibfnamefont{a.}~\bibnamefont{Pietzsch}},
  \bibinfo{author}{\bibfnamefont{M.}~\bibnamefont{Nagasono}},
  \bibinfo{author}{\bibfnamefont{F.}~\bibnamefont{Gel’mukhanov}},
  \bibinfo{author}{\bibfnamefont{L.}~\bibnamefont{Triguero}},
  \bibinfo{author}{\bibfnamefont{M.-N.} \bibnamefont{Piancastelli}},
  \bibinfo{author}{\bibfnamefont{W.}~\bibnamefont{Wurth}},
  \bibinfo{author}{\bibfnamefont{H.}~\bibnamefont{{\AA}gren}},
  \bibnamefont{et~al.}, \bibinfo{journal}{Phys. Rev. Lett.}
  \textbf{\bibinfo{volume}{95}}, \bibinfo{pages}{163002}
  (\bibinfo{year}{2005}).

\bibitem[{\citenamefont{Ljungberg et~al.}(2011)\citenamefont{Ljungberg,
  Pettersson, and Nilsson}}]{Ljungberg2011}
\bibinfo{author}{\bibfnamefont{M.~P.} \bibnamefont{Ljungberg}},
  \bibinfo{author}{\bibfnamefont{L.~G.~M.} \bibnamefont{Pettersson}},
  \bibnamefont{and} \bibinfo{author}{\bibfnamefont{A.}~\bibnamefont{Nilsson}},
  \bibinfo{journal}{J. Chem. Phys.} \textbf{\bibinfo{volume}{134}},
  \bibinfo{pages}{044513} (\bibinfo{year}{2011}).

\bibitem[{\citenamefont{Rubensson et~al.}(2013)\citenamefont{Rubensson,
  Hennies, and Pietzsch}}]{Rubensson2013}
\bibinfo{author}{\bibfnamefont{J.~E.} \bibnamefont{Rubensson}},
  \bibinfo{author}{\bibfnamefont{F.}~\bibnamefont{Hennies}}, \bibnamefont{and}
  \bibinfo{author}{\bibfnamefont{A.}~\bibnamefont{Pietzsch}},
  \bibinfo{journal}{J. Electron Spectros. Relat. Phenomena}
  \textbf{\bibinfo{volume}{188}}, \bibinfo{pages}{79} (\bibinfo{year}{2013}).

\bibitem[{\citenamefont{Guillemin et~al.}(2013)\citenamefont{Guillemin,
  Carniato, Journel, Stolte, Marchenko, Khoury, Kawerk, Piancastelli, Hudson,
  Lindle et~al.}}]{Guillemin2013}
\bibinfo{author}{\bibfnamefont{R.}~\bibnamefont{Guillemin}},
  \bibinfo{author}{\bibfnamefont{S.}~\bibnamefont{Carniato}},
  \bibinfo{author}{\bibfnamefont{L.}~\bibnamefont{Journel}},
  \bibinfo{author}{\bibfnamefont{W.~C.} \bibnamefont{Stolte}},
  \bibinfo{author}{\bibfnamefont{T.}~\bibnamefont{Marchenko}},
  \bibinfo{author}{\bibfnamefont{L.~E.} \bibnamefont{Khoury}},
  \bibinfo{author}{\bibfnamefont{E.}~\bibnamefont{Kawerk}},
  \bibinfo{author}{\bibfnamefont{M.~N.} \bibnamefont{Piancastelli}},
  \bibinfo{author}{\bibfnamefont{A.~C.} \bibnamefont{Hudson}},
  \bibinfo{author}{\bibfnamefont{D.~W.} \bibnamefont{Lindle}},
  \bibnamefont{et~al.}, \bibinfo{journal}{J. Electron Spectros. Relat.
  Phenomena} \textbf{\bibinfo{volume}{188}}, \bibinfo{pages}{53}
  (\bibinfo{year}{2013}).

\bibitem[{\citenamefont{Dong et~al.}(2013)\citenamefont{Dong, Wang, Olmstead,
  Fettinger, Nix, Uchiyama, Tsutsui, Baron, Dowty, and Cramer}}]{Dong2013}
\bibinfo{author}{\bibfnamefont{W.}~\bibnamefont{Dong}},
  \bibinfo{author}{\bibfnamefont{H.}~\bibnamefont{Wang}},
  \bibinfo{author}{\bibfnamefont{M.~M.} \bibnamefont{Olmstead}},
  \bibinfo{author}{\bibfnamefont{J.~C.} \bibnamefont{Fettinger}},
  \bibinfo{author}{\bibfnamefont{J.}~\bibnamefont{Nix}},
  \bibinfo{author}{\bibfnamefont{H.}~\bibnamefont{Uchiyama}},
  \bibinfo{author}{\bibfnamefont{S.}~\bibnamefont{Tsutsui}},
  \bibinfo{author}{\bibfnamefont{A.~Q.~R.} \bibnamefont{Baron}},
  \bibinfo{author}{\bibfnamefont{E.}~\bibnamefont{Dowty}}, \bibnamefont{and}
  \bibinfo{author}{\bibfnamefont{S.~P.} \bibnamefont{Cramer}},
  \bibinfo{journal}{Inorg. Chem.} \textbf{\bibinfo{volume}{52}},
  \bibinfo{pages}{6767} (\bibinfo{year}{2013}).

\bibitem[{\citenamefont{Bohinc et~al.}(2013)\citenamefont{Bohinc, {\v Z}itnik,
  Bu{\v c}ar, Kav{\v c}i{\v c}, Journel, Guillemin, Marchenko, Simon, and
  Cao}}]{Bohinc2013}
\bibinfo{author}{\bibfnamefont{R.}~\bibnamefont{Bohinc}},
  \bibinfo{author}{\bibfnamefont{M.}~\bibnamefont{{\v Z}itnik}},
  \bibinfo{author}{\bibfnamefont{K.}~\bibnamefont{Bu{\v c}ar}},
  \bibinfo{author}{\bibfnamefont{M.}~\bibnamefont{Kav{\v c}i{\v c}}},
  \bibinfo{author}{\bibfnamefont{L.}~\bibnamefont{Journel}},
  \bibinfo{author}{\bibfnamefont{R.}~\bibnamefont{Guillemin}},
  \bibinfo{author}{\bibfnamefont{T.}~\bibnamefont{Marchenko}},
  \bibinfo{author}{\bibfnamefont{M.}~\bibnamefont{Simon}}, \bibnamefont{and}
  \bibinfo{author}{\bibfnamefont{W.}~\bibnamefont{Cao}}, \bibinfo{journal}{J.
  Chem. Phys.} \textbf{\bibinfo{volume}{139}}, \bibinfo{pages}{134302}
  (\bibinfo{year}{2013}).

\bibitem[{\citenamefont{Pietzsch et~al.}(2011)\citenamefont{Pietzsch, Sun,
  Hennies, Rinkevicius, Karlsson, Schmitt, Strocov, Andersson, Kennedy,
  Schlappa et~al.}}]{Pietzsch2011}
\bibinfo{author}{\bibfnamefont{A.}~\bibnamefont{Pietzsch}},
  \bibinfo{author}{\bibfnamefont{Y.-P.} \bibnamefont{Sun}},
  \bibinfo{author}{\bibfnamefont{F.}~\bibnamefont{Hennies}},
  \bibinfo{author}{\bibfnamefont{Z.}~\bibnamefont{Rinkevicius}},
  \bibinfo{author}{\bibfnamefont{H.~O.} \bibnamefont{Karlsson}},
  \bibinfo{author}{\bibfnamefont{T.}~\bibnamefont{Schmitt}},
  \bibinfo{author}{\bibfnamefont{V.~N.} \bibnamefont{Strocov}},
  \bibinfo{author}{\bibfnamefont{J.}~\bibnamefont{Andersson}},
  \bibinfo{author}{\bibfnamefont{B.}~\bibnamefont{Kennedy}},
  \bibinfo{author}{\bibfnamefont{J.}~\bibnamefont{Schlappa}},
  \bibnamefont{et~al.}, \bibinfo{journal}{Phys. Rev. Lett.}
  \textbf{\bibinfo{volume}{106}}, \bibinfo{pages}{153004}
  (\bibinfo{year}{2011}).

\bibitem[{\citenamefont{Hennies et~al.}(2010)\citenamefont{Hennies, Pietzsch,
  Berglund, F{\"{o}}hlisch, Schmitt, Strocov, Karlsson, Andersson, and
  Rubensson}}]{Hennies2010}
\bibinfo{author}{\bibfnamefont{F.}~\bibnamefont{Hennies}},
  \bibinfo{author}{\bibfnamefont{A.}~\bibnamefont{Pietzsch}},
  \bibinfo{author}{\bibfnamefont{M.}~\bibnamefont{Berglund}},
  \bibinfo{author}{\bibfnamefont{A.}~\bibnamefont{F{\"{o}}hlisch}},
  \bibinfo{author}{\bibfnamefont{T.}~\bibnamefont{Schmitt}},
  \bibinfo{author}{\bibfnamefont{V.}~\bibnamefont{Strocov}},
  \bibinfo{author}{\bibfnamefont{H.~O.} \bibnamefont{Karlsson}},
  \bibinfo{author}{\bibfnamefont{J.}~\bibnamefont{Andersson}},
  \bibnamefont{and} \bibinfo{author}{\bibfnamefont{J.-E.}
  \bibnamefont{Rubensson}}, \bibinfo{journal}{Phys. Rev. Lett.}
  \textbf{\bibinfo{volume}{104}}, \bibinfo{pages}{193002}
  (\bibinfo{year}{2010}).

\bibitem[{\citenamefont{Lange and Aziz}(2013)}]{Lange2013}
\bibinfo{author}{\bibfnamefont{K.~M.} \bibnamefont{Lange}} \bibnamefont{and}
  \bibinfo{author}{\bibfnamefont{E.~F.} \bibnamefont{Aziz}},
  \bibinfo{journal}{Chem. Asian J.} \textbf{\bibinfo{volume}{8}},
  \bibinfo{pages}{318} (\bibinfo{year}{2013}).

\bibitem[{\citenamefont{Schreck et~al.}(2014)\citenamefont{Schreck, Pietzsch,
  Kunnus, Kennedy, Quevedo, Miedema, Wernet, and F{\"o}hlisch}}]{Schreck2014}
\bibinfo{author}{\bibfnamefont{S.}~\bibnamefont{Schreck}},
  \bibinfo{author}{\bibfnamefont{A.}~\bibnamefont{Pietzsch}},
  \bibinfo{author}{\bibfnamefont{K.}~\bibnamefont{Kunnus}},
  \bibinfo{author}{\bibfnamefont{B.}~\bibnamefont{Kennedy}},
  \bibinfo{author}{\bibfnamefont{W.}~\bibnamefont{Quevedo}},
  \bibinfo{author}{\bibfnamefont{P.~S.} \bibnamefont{Miedema}},
  \bibinfo{author}{\bibfnamefont{P.}~\bibnamefont{Wernet}}, \bibnamefont{and}
  \bibinfo{author}{\bibfnamefont{A.}~\bibnamefont{F{\"o}hlisch}},
  \bibinfo{journal}{Struct. Dyn.} \textbf{\bibinfo{volume}{1}}
  (\bibinfo{year}{2014}).

\bibitem[{\citenamefont{Fransson et~al.}(2016)\citenamefont{Fransson, Harada,
  Kosugi, Besley, Winter, Rehr, Pettersson, and Nilsson}}]{Fransson-CR-2016}
\bibinfo{author}{\bibfnamefont{T.}~\bibnamefont{Fransson}},
  \bibinfo{author}{\bibfnamefont{Y.}~\bibnamefont{Harada}},
  \bibinfo{author}{\bibfnamefont{N.}~\bibnamefont{Kosugi}},
  \bibinfo{author}{\bibfnamefont{N.~A.} \bibnamefont{Besley}},
  \bibinfo{author}{\bibfnamefont{B.}~\bibnamefont{Winter}},
  \bibinfo{author}{\bibfnamefont{J.~J.} \bibnamefont{Rehr}},
  \bibinfo{author}{\bibfnamefont{L.~G.~M.} \bibnamefont{Pettersson}},
  \bibnamefont{and} \bibinfo{author}{\bibfnamefont{A.}~\bibnamefont{Nilsson}},
  \bibinfo{journal}{Chemical Reviews} \textbf{\bibinfo{volume}{116}},
  \bibinfo{pages}{7551} (\bibinfo{year}{2016}).

\bibitem[{\citenamefont{Sellberg et~al.}(2015)\citenamefont{Sellberg, Mcqueen,
  Laksmono, Schreck, Beye, Deponte, Kennedy, Nordlund, Sierra, Schlesinger
  et~al.}}]{Sellberg2015}
\bibinfo{author}{\bibfnamefont{J.~A.} \bibnamefont{Sellberg}},
  \bibinfo{author}{\bibfnamefont{T.~A.} \bibnamefont{Mcqueen}},
  \bibinfo{author}{\bibfnamefont{H.}~\bibnamefont{Laksmono}},
  \bibinfo{author}{\bibfnamefont{S.}~\bibnamefont{Schreck}},
  \bibinfo{author}{\bibfnamefont{M.}~\bibnamefont{Beye}},
  \bibinfo{author}{\bibfnamefont{D.~P.} \bibnamefont{Deponte}},
  \bibinfo{author}{\bibfnamefont{B.}~\bibnamefont{Kennedy}},
  \bibinfo{author}{\bibfnamefont{D.}~\bibnamefont{Nordlund}},
  \bibinfo{author}{\bibfnamefont{R.~G.} \bibnamefont{Sierra}},
  \bibinfo{author}{\bibfnamefont{D.}~\bibnamefont{Schlesinger}},
  \bibnamefont{et~al.}, \bibinfo{journal}{J. Chem. Phys.}
  (\bibinfo{year}{2015}).

\bibitem[{\citenamefont{Mukamel}(1995)}]{Mukamel-Book}
\bibinfo{author}{\bibfnamefont{S.}~\bibnamefont{Mukamel}},
  \emph{\bibinfo{title}{{Principles of Nonlinear Optical Spectroscopy}}}
  (\bibinfo{publisher}{Oxford University Press, Oxford}, \bibinfo{year}{1995}).

\bibitem[{\citenamefont{May and K\"{u}hn}(2011)}]{Kuehn-Book}
\bibinfo{author}{\bibfnamefont{V.}~\bibnamefont{May}} \bibnamefont{and}
  \bibinfo{author}{\bibfnamefont{O.}~\bibnamefont{K\"{u}hn}},
  \emph{\bibinfo{title}{{Charge and Energy Transfer Dynamics in Molecular
  Systems}}} (\bibinfo{publisher}{Wiley-VCH}, \bibinfo{year}{2011}), ISBN
  \bibinfo{isbn}{978-3-527-40732-3}.

\bibitem[{\citenamefont{Marx and Hutter}(2009)}]{marx2009}
\bibinfo{author}{\bibfnamefont{D.}~\bibnamefont{Marx}} \bibnamefont{and}
  \bibinfo{author}{\bibfnamefont{J.}~\bibnamefont{Hutter}},
  \emph{\bibinfo{title}{Ab initio molecular dynamics: basic theory and advanced
  methods}} (\bibinfo{publisher}{Cambridge University Press},
  \bibinfo{year}{2009}).

\bibitem[{\citenamefont{Ivanov et~al.}(2013)\citenamefont{Ivanov, Witt, and
  Marx}}]{Ivanov-PCCP-2013}
\bibinfo{author}{\bibfnamefont{S.~D.} \bibnamefont{Ivanov}},
  \bibinfo{author}{\bibfnamefont{A.}~\bibnamefont{Witt}}, \bibnamefont{and}
  \bibinfo{author}{\bibfnamefont{D.}~\bibnamefont{Marx}},
  \bibinfo{journal}{Phys. Chem. Chem. Phys.} \textbf{\bibinfo{volume}{15}},
  \bibinfo{pages}{10270} (\bibinfo{year}{2013}).

\bibitem[{\citenamefont{Sun et~al.}(2011)\citenamefont{Sun, Hennies, Pietzsch,
  Kennedy, Schmitt, Strocov, Andersson, Berglund, Rubensson, Aidas
  et~al.}}]{Sun2011}
\bibinfo{author}{\bibfnamefont{Y.-P.} \bibnamefont{Sun}},
  \bibinfo{author}{\bibfnamefont{F.}~\bibnamefont{Hennies}},
  \bibinfo{author}{\bibfnamefont{A.}~\bibnamefont{Pietzsch}},
  \bibinfo{author}{\bibfnamefont{B.}~\bibnamefont{Kennedy}},
  \bibinfo{author}{\bibfnamefont{T.}~\bibnamefont{Schmitt}},
  \bibinfo{author}{\bibfnamefont{V.~N.} \bibnamefont{Strocov}},
  \bibinfo{author}{\bibfnamefont{J.}~\bibnamefont{Andersson}},
  \bibinfo{author}{\bibfnamefont{M.}~\bibnamefont{Berglund}},
  \bibinfo{author}{\bibfnamefont{J.-E.} \bibnamefont{Rubensson}},
  \bibinfo{author}{\bibfnamefont{K.}~\bibnamefont{Aidas}},
  \bibnamefont{et~al.}, \bibinfo{journal}{Phys. Rev. B}
  \textbf{\bibinfo{volume}{84}}, \bibinfo{pages}{132202}
  (\bibinfo{year}{2011}).

\bibitem[{\citenamefont{Jena et~al.}(2015)\citenamefont{Jena, Josefsson,
  Eriksson, Hagfeldt, Siegbahn, Bj{\"{o}}rneholm, Rensmo, and
  Odelius}}]{Jena2015}
\bibinfo{author}{\bibfnamefont{N.~K.} \bibnamefont{Jena}},
  \bibinfo{author}{\bibfnamefont{I.}~\bibnamefont{Josefsson}},
  \bibinfo{author}{\bibfnamefont{S.~K.} \bibnamefont{Eriksson}},
  \bibinfo{author}{\bibfnamefont{A.}~\bibnamefont{Hagfeldt}},
  \bibinfo{author}{\bibfnamefont{H.}~\bibnamefont{Siegbahn}},
  \bibinfo{author}{\bibfnamefont{O.}~\bibnamefont{Bj{\"{o}}rneholm}},
  \bibinfo{author}{\bibfnamefont{H.}~\bibnamefont{Rensmo}}, \bibnamefont{and}
  \bibinfo{author}{\bibfnamefont{M.}~\bibnamefont{Odelius}},
  \bibinfo{journal}{Chem. - A Eur. J.} \textbf{\bibinfo{volume}{21}},
  \bibinfo{pages}{4049} (\bibinfo{year}{2015}).

\bibitem[{\citenamefont{Weinhardt et~al.}(2015)\citenamefont{Weinhardt, Ertan,
  Iannuzzi, Weigand, Fuchs, B{\"{a}}r, Blum, Denlinger, Yang, Umbach
  et~al.}}]{Weinhardt2015}
\bibinfo{author}{\bibfnamefont{L.}~\bibnamefont{Weinhardt}},
  \bibinfo{author}{\bibfnamefont{E.}~\bibnamefont{Ertan}},
  \bibinfo{author}{\bibfnamefont{M.}~\bibnamefont{Iannuzzi}},
  \bibinfo{author}{\bibfnamefont{M.}~\bibnamefont{Weigand}},
  \bibinfo{author}{\bibfnamefont{O.}~\bibnamefont{Fuchs}},
  \bibinfo{author}{\bibfnamefont{M.}~\bibnamefont{B{\"{a}}r}},
  \bibinfo{author}{\bibfnamefont{M.}~\bibnamefont{Blum}},
  \bibinfo{author}{\bibfnamefont{J.~D.} \bibnamefont{Denlinger}},
  \bibinfo{author}{\bibfnamefont{W.}~\bibnamefont{Yang}},
  \bibinfo{author}{\bibfnamefont{E.}~\bibnamefont{Umbach}},
  \bibnamefont{et~al.}, \bibinfo{journal}{Phys. Chem. Chem. Phys.}
  \textbf{\bibinfo{volume}{17}}, \bibinfo{pages}{27145} (\bibinfo{year}{2015}).

\bibitem[{\citenamefont{Leetmaa et~al.}(2010)\citenamefont{Leetmaa, Ljungberg,
  Lyubartsev, Nilsson, and Pettersson}}]{Leetmaa2010}
\bibinfo{author}{\bibfnamefont{M.}~\bibnamefont{Leetmaa}},
  \bibinfo{author}{\bibfnamefont{M.}~\bibnamefont{Ljungberg}},
  \bibinfo{author}{\bibfnamefont{A.}~\bibnamefont{Lyubartsev}},
  \bibinfo{author}{\bibfnamefont{A.}~\bibnamefont{Nilsson}}, \bibnamefont{and}
  \bibinfo{author}{\bibfnamefont{L.}~\bibnamefont{Pettersson}},
  \bibinfo{journal}{J. Electron Spectros. Relat. Phenomena}
  \textbf{\bibinfo{volume}{177}}, \bibinfo{pages}{135} (\bibinfo{year}{2010}).

\bibitem[{\citenamefont{Heller}(1978)}]{Heller1978}
\bibinfo{author}{\bibfnamefont{E.~J.} \bibnamefont{Heller}},
  \bibinfo{journal}{J. Chem. Phys.} \textbf{\bibinfo{volume}{68}},
  \bibinfo{pages}{3891} (\bibinfo{year}{1978}).

\bibitem[{\citenamefont{Lee and Heller}(1979)}]{Heller-JCP-1979}
\bibinfo{author}{\bibfnamefont{S.}~\bibnamefont{Lee}} \bibnamefont{and}
  \bibinfo{author}{\bibfnamefont{E.~J.} \bibnamefont{Heller}},
  \bibinfo{journal}{J. Chem. Phys.} \textbf{\bibinfo{volume}{71}},
  \bibinfo{pages}{4777} (\bibinfo{year}{1979}),
  \urlprefix\url{http://scitation.aip.org/content/aip/journal/jcp/71/12/10.1063/1.438316}.

\bibitem[{\citenamefont{Lawrence and Skinner}(2002)}]{lawrence2002}
\bibinfo{author}{\bibfnamefont{C.}~\bibnamefont{Lawrence}} \bibnamefont{and}
  \bibinfo{author}{\bibfnamefont{J.}~\bibnamefont{Skinner}},
  \bibinfo{journal}{J. Chem. Phys.} \textbf{\bibinfo{volume}{117}},
  \bibinfo{pages}{8847} (\bibinfo{year}{2002}).

\bibitem[{\citenamefont{Harder et~al.}(2005)\citenamefont{Harder, Eaves,
  Tokmakoff, and Berne}}]{harder2005}
\bibinfo{author}{\bibfnamefont{E.}~\bibnamefont{Harder}},
  \bibinfo{author}{\bibfnamefont{J.~D.} \bibnamefont{Eaves}},
  \bibinfo{author}{\bibfnamefont{A.}~\bibnamefont{Tokmakoff}},
  \bibnamefont{and} \bibinfo{author}{\bibfnamefont{B.}~\bibnamefont{Berne}},
  \bibinfo{journal}{Proc. Nat. Acad. Sci.} \textbf{\bibinfo{volume}{102}},
  \bibinfo{pages}{11611} (\bibinfo{year}{2005}).

\bibitem[{\citenamefont{Lange et~al.}(2012)\citenamefont{Lange, Kothe, and
  Aziz}}]{Lange2012}
\bibinfo{author}{\bibfnamefont{K.~M.} \bibnamefont{Lange}},
  \bibinfo{author}{\bibfnamefont{A.}~\bibnamefont{Kothe}}, \bibnamefont{and}
  \bibinfo{author}{\bibfnamefont{E.~F.} \bibnamefont{Aziz}},
  \bibinfo{journal}{Phys. Chem. Chem. Phys.} \textbf{\bibinfo{volume}{14}},
  \bibinfo{pages}{5331} (\bibinfo{year}{2012}).

\bibitem[{\citenamefont{Hess et~al.}(2008)\citenamefont{Hess, Kutzner, van~der
  Spoel, and Lindahl}}]{GROMACS}
\bibinfo{author}{\bibfnamefont{B.}~\bibnamefont{Hess}},
  \bibinfo{author}{\bibfnamefont{C.}~\bibnamefont{Kutzner}},
  \bibinfo{author}{\bibfnamefont{D.}~\bibnamefont{van~der Spoel}},
  \bibnamefont{and} \bibinfo{author}{\bibfnamefont{E.}~\bibnamefont{Lindahl}},
  \bibinfo{journal}{J. Chem. Theory Comput.} \textbf{\bibinfo{volume}{4}},
  \bibinfo{pages}{435} (\bibinfo{year}{2008}).

\bibitem[{\citenamefont{Paesani et~al.}(2006)\citenamefont{Paesani, Zhang,
  Case, Cheatham, and Voth}}]{Paesani-JCP-2006}
\bibinfo{author}{\bibfnamefont{F.}~\bibnamefont{Paesani}},
  \bibinfo{author}{\bibfnamefont{W.}~\bibnamefont{Zhang}},
  \bibinfo{author}{\bibfnamefont{D.~A.} \bibnamefont{Case}},
  \bibinfo{author}{\bibfnamefont{T.~E.} \bibnamefont{Cheatham}},
  \bibnamefont{and} \bibinfo{author}{\bibfnamefont{G.~A.} \bibnamefont{Voth}},
  \bibinfo{journal}{J. Chem. Phys.} \textbf{\bibinfo{volume}{125}},
  \bibinfo{pages}{184507} (\bibinfo{year}{2006}).

\bibitem[{\citenamefont{Perdew et~al.}(1996)\citenamefont{Perdew, Burke, and
  Ernzerhof}}]{PBE_1996}
\bibinfo{author}{\bibfnamefont{J.~P.} \bibnamefont{Perdew}},
  \bibinfo{author}{\bibfnamefont{K.}~\bibnamefont{Burke}}, \bibnamefont{and}
  \bibinfo{author}{\bibfnamefont{M.}~\bibnamefont{Ernzerhof}},
  \bibinfo{journal}{Phys. Rev. Lett.} \textbf{\bibinfo{volume}{77}},
  \bibinfo{pages}{3865} (\bibinfo{year}{1996}).

\bibitem[{\citenamefont{Neese}(2012)}]{orca_2012}
\bibinfo{author}{\bibfnamefont{F.}~\bibnamefont{Neese}},
  \bibinfo{journal}{Wiley Interdiscip. Rev. Comput. Mol. Sci.}
  \textbf{\bibinfo{volume}{2}}, \bibinfo{pages}{73} (\bibinfo{year}{2012}).

\bibitem[{\citenamefont{Weigend and Ahlrichs}(2005)}]{def2_2005}
\bibinfo{author}{\bibfnamefont{F.}~\bibnamefont{Weigend}} \bibnamefont{and}
  \bibinfo{author}{\bibfnamefont{R.}~\bibnamefont{Ahlrichs}},
  \bibinfo{journal}{Phys. Chem. Chem. Phys.} \textbf{\bibinfo{volume}{7}},
  \bibinfo{pages}{3297} (\bibinfo{year}{2005}).

\bibitem[{\citenamefont{N\"aslund et~al.}(2003)\citenamefont{N\"aslund,
  Cavalleri, Ogasawara, Nilsson, Pettersson, Wernet, Edwards, Sandstr\"om, and
  Myneni}}]{naeslund_JPCA_2003}
\bibinfo{author}{\bibfnamefont{L.-{\AA}.} \bibnamefont{N\"aslund}},
  \bibinfo{author}{\bibfnamefont{M.}~\bibnamefont{Cavalleri}},
  \bibinfo{author}{\bibfnamefont{H.}~\bibnamefont{Ogasawara}},
  \bibinfo{author}{\bibfnamefont{A.}~\bibnamefont{Nilsson}},
  \bibinfo{author}{\bibfnamefont{L.~G.~M.} \bibnamefont{Pettersson}},
  \bibinfo{author}{\bibfnamefont{P.}~\bibnamefont{Wernet}},
  \bibinfo{author}{\bibfnamefont{D.~C.} \bibnamefont{Edwards}},
  \bibinfo{author}{\bibfnamefont{M.}~\bibnamefont{Sandstr\"om}},
  \bibnamefont{and} \bibinfo{author}{\bibfnamefont{S.}~\bibnamefont{Myneni}},
  \bibinfo{journal}{The Journal of Physical Chemistry A}
  \textbf{\bibinfo{volume}{107}}, \bibinfo{pages}{6869} (\bibinfo{year}{2003}).

\bibitem[{\citenamefont{Lee et~al.}(2010)\citenamefont{Lee, Petrenko, Bergmann,
  Neese, and DeBeer}}]{lee_JACS_2010}
\bibinfo{author}{\bibfnamefont{N.}~\bibnamefont{Lee}},
  \bibinfo{author}{\bibfnamefont{T.}~\bibnamefont{Petrenko}},
  \bibinfo{author}{\bibfnamefont{U.}~\bibnamefont{Bergmann}},
  \bibinfo{author}{\bibfnamefont{F.}~\bibnamefont{Neese}}, \bibnamefont{and}
  \bibinfo{author}{\bibfnamefont{S.}~\bibnamefont{DeBeer}},
  \bibinfo{journal}{J. Am. Chem. Soc.} \textbf{\bibinfo{volume}{132}},
  \bibinfo{pages}{9715} (\bibinfo{year}{2010}).

\bibitem[{\citenamefont{Lassalle-Kaiser
  et~al.}(2013)\citenamefont{Lassalle-Kaiser, Boron, Krewald, Kern, Beckwith,
  Delgado-Jaime, Schroeder, Alonso-Mori, Nordlund, Weng
  et~al.}}]{Lassalle-Kaiser2013}
\bibinfo{author}{\bibfnamefont{B.}~\bibnamefont{Lassalle-Kaiser}},
  \bibinfo{author}{\bibfnamefont{T.~T.} \bibnamefont{Boron}},
  \bibinfo{author}{\bibfnamefont{V.}~\bibnamefont{Krewald}},
  \bibinfo{author}{\bibfnamefont{J.}~\bibnamefont{Kern}},
  \bibinfo{author}{\bibfnamefont{M.~a.} \bibnamefont{Beckwith}},
  \bibinfo{author}{\bibfnamefont{M.~U.} \bibnamefont{Delgado-Jaime}},
  \bibinfo{author}{\bibfnamefont{H.}~\bibnamefont{Schroeder}},
  \bibinfo{author}{\bibfnamefont{R.}~\bibnamefont{Alonso-Mori}},
  \bibinfo{author}{\bibfnamefont{D.}~\bibnamefont{Nordlund}},
  \bibinfo{author}{\bibfnamefont{T.~C.} \bibnamefont{Weng}},
  \bibnamefont{et~al.}, \bibinfo{journal}{Inorg. Chem.}
  \textbf{\bibinfo{volume}{52}}, \bibinfo{pages}{12915} (\bibinfo{year}{2013}).

\bibitem[{\citenamefont{Pollock et~al.}(2013)\citenamefont{Pollock, Grubel,
  Holland, and Debeer}}]{Pollock2013}
\bibinfo{author}{\bibfnamefont{C.~J.} \bibnamefont{Pollock}},
  \bibinfo{author}{\bibfnamefont{K.}~\bibnamefont{Grubel}},
  \bibinfo{author}{\bibfnamefont{P.~L.} \bibnamefont{Holland}},
  \bibnamefont{and} \bibinfo{author}{\bibfnamefont{S.}~\bibnamefont{Debeer}},
  \bibinfo{journal}{J. Am. Chem. Soc.} \textbf{\bibinfo{volume}{135}},
  \bibinfo{pages}{11803} (\bibinfo{year}{2013}).

\bibitem[{\citenamefont{Harada et~al.}(2013)\citenamefont{Harada, Tokushima,
  Horikawa, Takahashi, Niwa, Kobayashi, Oshima, Senba, Ohashi, Wikfeldt
  et~al.}}]{Harada-PRL-2013}
\bibinfo{author}{\bibfnamefont{Y.}~\bibnamefont{Harada}},
  \bibinfo{author}{\bibfnamefont{T.}~\bibnamefont{Tokushima}},
  \bibinfo{author}{\bibfnamefont{Y.}~\bibnamefont{Horikawa}},
  \bibinfo{author}{\bibfnamefont{O.}~\bibnamefont{Takahashi}},
  \bibinfo{author}{\bibfnamefont{H.}~\bibnamefont{Niwa}},
  \bibinfo{author}{\bibfnamefont{M.}~\bibnamefont{Kobayashi}},
  \bibinfo{author}{\bibfnamefont{M.}~\bibnamefont{Oshima}},
  \bibinfo{author}{\bibfnamefont{Y.}~\bibnamefont{Senba}},
  \bibinfo{author}{\bibfnamefont{H.}~\bibnamefont{Ohashi}},
  \bibinfo{author}{\bibfnamefont{K.~T.} \bibnamefont{Wikfeldt}},
  \bibnamefont{et~al.}, \bibinfo{journal}{Phys. Rev. Lett.}
  \textbf{\bibinfo{volume}{111}}, \bibinfo{pages}{1} (\bibinfo{year}{2013}).

\bibitem[{\citenamefont{Biggs et~al.}(2012)\citenamefont{Biggs, Zhang, Healion,
  and Mukamel}}]{Biggs2012}
\bibinfo{author}{\bibfnamefont{J.~D.} \bibnamefont{Biggs}},
  \bibinfo{author}{\bibfnamefont{Y.}~\bibnamefont{Zhang}},
  \bibinfo{author}{\bibfnamefont{D.}~\bibnamefont{Healion}}, \bibnamefont{and}
  \bibinfo{author}{\bibfnamefont{S.}~\bibnamefont{Mukamel}},
  \bibinfo{journal}{J. Chem. Phys.} \textbf{\bibinfo{volume}{136}},
  \bibinfo{pages}{174117} (\bibinfo{year}{2012}).

\bibitem[{\citenamefont{Mukamel et~al.}(2013)\citenamefont{Mukamel, Healion,
  Zhang, and Biggs}}]{Mukamel2013}
\bibinfo{author}{\bibfnamefont{S.}~\bibnamefont{Mukamel}},
  \bibinfo{author}{\bibfnamefont{D.}~\bibnamefont{Healion}},
  \bibinfo{author}{\bibfnamefont{Y.}~\bibnamefont{Zhang}}, \bibnamefont{and}
  \bibinfo{author}{\bibfnamefont{J.~D.} \bibnamefont{Biggs}},
  \bibinfo{journal}{Annu. Rev. Phys. Chem.} \textbf{\bibinfo{volume}{64}},
  \bibinfo{pages}{101} (\bibinfo{year}{2013}).

\end{thebibliography}

\end{document}